\documentclass[journal]{IEEEtran}
\usepackage{amsfonts}
\usepackage{amssymb,bm,graphicx,subfigure}

\ifCLASSINFOpdf
\else
\fi
\usepackage[cmex10]{amsmath}
\begin{document}
%
\title{Line Spectral Estimation Based on Compressed Sensing with Deterministic Sub-Nyquist Sampling}
%
%
%
\author{Shan~Huang,~
        Hong~Sun,~
        Haijian~Zhang,~
        and Lei~Yu
\thanks{The authors are with Signal Processing Laboratory, School of Electronic Information, Wuhan University, Wuhan 430072, China (e-mail: staronice@whu.edu.cn; hongsun@whu.edu.cn; haijian.zhang@whu.edu.cn; ly.wd@whu.edu.cn).}
\thanks{}
}

%
%

\markboth{Journal of \LaTeX\ Class Files,~Vol.~11, No.~4, December~2012}%
{Shell \MakeLowercase{\textit{et al.}}: Bare Demo of IEEEtran.cls for Journals}
%



\maketitle

\begin{abstract}
As an alternative to the traditional sampling theory, compressed sensing allows acquiring much smaller amount of data, still estimating the spectra of frequency-sparse signals accurately. However, compressed sensing usually requires random sampling in data acquisition, which is difficult to implement in hardware. In this paper, we propose a deterministic and simple sampling scheme, that is, sampling at three sub-Nyquist rates which have coprime undersampled ratios. This sampling method turns out to be valid through numerical experiments. A complex-valued multitask algorithm based on variational Bayesian inference is proposed to estimate the spectra of frequency-sparse signals after sampling. Simulations show that this method is feasible and robust at quite low sampling rates.
\end{abstract}


%
\IEEEpeerreviewmaketitle

\section{Introduction}
%
%
%
%
\IEEEPARstart{L}{ine} spectral estimation has numerous applications in sonar, radar, underwater surveillance, communications, geophysical exploration, speech analysis, nuclear physics and other fields. In general, the sampling rate of the signal is required to be higher than twice the highest frequency (i.e., Nyquist rate). In some applications, high-speed analog-to-digital converters that increase the sampling rate or density are very expensive. Emerging compressed sensing (CS) goes against the common knowledge in data acquisition. CS theory asserts that one can recover certain signals and images from far fewer samples or measurements than traditional methods use \cite{donoho2006compressed}.

Many researchers have utilized CS to estimate the spectra of frequency-sparse signals \cite{zhu2011sparsity}\cite{duarte2013spectral}\cite{chen2014robust}. A source localization method based on a sparse representation of sensor measurements with an overcomplete basis was proposed in \cite{malioutov2005sparse}. The authors in \cite{bourguignon2007sparsity} addressed the problem of estimating spectral lines from irregularly sampled data within the framework of sparse representations. The uniqueness conditions of the sparse solution with different patterns of samples were analyzed. In \cite{chi2011sensitivity}, the effect of "basis mismatch" caused by grid discretization was analyzed. To deal with basis mismatch, some articles used grid refinement to approximate the true grid \cite{yang2012robustly}\cite{hu2012compressed}\cite{fang2014super}. The atomic norm-based methods make line spectral estimation cast into a convex semidefinite program optimization, which deals with continuous-valued frequencies and completely eliminates basis mismatch \cite{candes2014towards}\cite{tang2013compressed}\cite{yang2015gridless}. However, these methods usually require random sampling, which is difficult or complicated to implement.

In this paper, we focus on line spectral estimation with deterministic sub-Nyquist sampling. The union of three series of undersampled samples at coprime ratios is enough to estimate the spectra of frequency-sparse signals. Then an algorithm based on variational Bayesian inference is employed to connect the samples. This method may be realized through three undersampled channels, the hardware is convenient to implement. The paper is organized as follows: Section II gives the smapling strategy. Section III demonstrates our algorithm. Simulation results are shown in Section IV. The last section draws conclusions.

\section{Sampling Strategy }
Consider the line spectral estimation problem where the observed signal is a summation of $K$ complex sinusoids:
\begin{equation}\label{eq1}
  y(m) = \sum\limits_{k = 1}^K {{c_k}{e^{ j{\omega _k}m}}},
\end{equation}
where $j=\sqrt{-1}$, ${\omega _k} \in \left[ {0,2\pi } \right)$ and $c_k$ denote the angular frequency and the complex amplitude of the $k$-th component, respectively. When $m=1,2,\cdots,M$, it implies normal sampling, which is studied in conventional methods such as MUSIC \cite{schmidt1986multiple}. In the methods based on CS, $m$ is selected at random from the index set $\left[ N \right] \buildrel \Delta \over = \left\{ {1,2, \cdots ,N} \right\}$. However, this pattern of sampling often leads to complex hardware. For example, a new type of data acquisition system called a random demodulator is studied to ensure the randomness of sampling in \cite{tropp2010beyond}.

\begin{figure}[!htbp]
\centering
\includegraphics[scale=0.8]{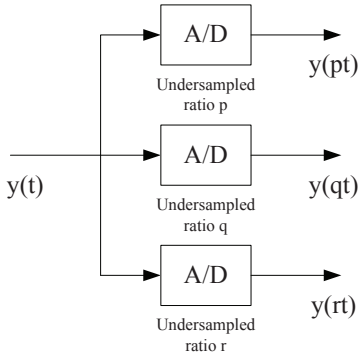}
\centering
\caption{ The process diagram of the proposed sampling scheme. } \label{fig1}
\end{figure}

The proposed deterministic scheme is to sample at three coprime undersampled ratios $p,q,r$, in other words, we need the samples with indices
\begin{equation}\label{eq2}
 \mathcal{I }= \left\{ {p,2p, \cdots } \right\} \cup \left\{ {q,2q, \cdots } \right\} \cup \left\{ {r,2r, \cdots } \right\}.
\end{equation}
It is worth mentioning that sampling at two coprime undersampled ratios sometimes also yields correct results but three rates guarantee a high probability of success. The process diagram of sampling is shown in Fig.~\ref{fig1}.

After sampling, the samples are in chronological order, we select consecutive $M$ samples with indices $t_1,t_2,\cdots,t_M$ to constitute a column vector
\begin{equation}\label{eq3}
  \bm{y} = {\left[ {\begin{array}{*{20}{c}}{y({t_1})}&{y({t_2})}& \cdots &{y({t_M})}\end{array}} \right]^T},
\end{equation}
where $[\ast]^{T}$ denotes the transpose operation. Assume that the frequencies are aligned with a uniform grid, i.e.,
\begin{equation}\label{eq4}
{\omega _n} = {{2\pi n} \mathord{\left/
 {\vphantom {{2\pi n} N}} \right.
 \kern-\nulldelimiterspace} N},~n = 1,2, \cdots ,N.
\end{equation}
The observation model can be written more compactly as
\begin{equation}\label{eq5}
  \bm y = \bm \Phi(\omega )\bm s,
\end{equation}
where $\bm \Phi(\omega ) = \left[ {\begin{array}{*{20}{c}}
{\bm \phi({\omega _1})}&{\bm \phi({\omega _2})}& \cdots &{\bm \phi({\omega _N})}
\end{array}} \right]$, $\bm {\phi}(\omega ) = {\left[ {\begin{array}{*{20}{c}}
{{e^{j\omega {t_1}}}}& \cdots &{{e^{j\omega {t_M}}}}
\end{array}} \right]^T}$ and $\bm s = {\left[ {\begin{array}{*{20}{c}}
{{\tilde{c}_1}}& \cdots &{{\tilde{c}_N}}
\end{array}} \right]^T}$ is a $K$-sparse vector. In general, $M<N$ is set and (\ref{eq5}) is solved as a problem of sparse recovery. However, the property of $\bm \Phi$ as a CS matrix is difficult to certify in theory. To improve the probability of success, we utilize more samples to form multiple tasks and synthesize the effects of these tasks, namely
\begin{equation}\label{eq6}
  {\bm{y}_l} = {\bm{\Phi}_l}{\bm{s}_{l}},~l = 1,2, \cdots ,L,
\end{equation}
where
\begin{equation}\label{eq7}
{\bm \Phi _l} = \left[ {\begin{array}{*{20}{c}}
1& \cdots &1\\
{{e^{j{\omega _1}\left( {{t_{l + 1}} - {t_l}} \right)}}}& \cdots &{{e^{j{\omega _N}\left( {{t_{l + 1}} - {t_l}} \right)}}}\\
 \vdots & \ddots & \vdots \\
{{e^{j{\omega _1}\left( {{t_{l + M - 1}} - {t_l}} \right)}}}& \cdots &{{e^{j{\omega _N}\left( {{t_{l + M - 1}} - {t_l}} \right)}}}
\end{array}} \right],
\end{equation}
${\bm y_l} = {\left[ {\begin{array}{*{20}{c}}
{y({t_l})}&{y({t_{l + 1}})}& \cdots &{y({t_{l + M - 1}})}
\end{array}} \right]^T}$ and ${\bm s_l} = {\left[ {\begin{array}{*{20}{c}}
{{\tilde{c}_1}{e^{j{\omega _1}{t_l}}}}& \cdots &{{\tilde{c}_N}{e^{j{\omega _N}{t_l}}}}
\end{array}} \right]^T}$. The total number of samples is $L+M-1$. Note that all of $\bm s_{l}$ share the same sparsity profile and $\bm \Phi_{l}$ repeat after a certain period. $M$ is expected to be as large as possible, but an appropriate value of $M$ must ensure $\bm \Phi_{l}$ not to contain duplicate rows. The joint estimation can achieve satisfactory results as shown in Section IV, even though $\bm \Phi_l$ may not have good property.

The most widely used criterion to evaluate the property of a CS matrix is \emph{restricted isometry property} (RIP). The CS matrix $\bm{\Phi}$ has the ($\emph{k},\delta$)-RIP if
\begin{equation}\label{eq7p}
\left( {1 - {\delta}} \right)\left\| \bm x \right\|_2^2 \le \left\| {{\bm{\Phi }}\bm x} \right\|_2^2 \le \left( {1 + {\delta}} \right)\left\| \bm x \right\|_2^2
\end{equation}
holds for all \emph{k}-sparse vectors $\bm x$, $\|\bm x\|_{2}$ denotes $\ell_{2}$-norm of $\bm x$ \cite{candes2008introduction}. The smallest $\delta$ for ($\emph{k},\delta$)-RIP is the restricted isometry constant (RIC) $\delta_{k}$. A small $\delta_{k}$ implies good performance when recovering a $k$-sparse signal. Let $\bm{\Phi}$ be a matrix with $\ell_{2}$-normalized columns $\bm \varphi_{1},\bm \varphi_{2},\cdots,\bm \varphi_{N}$, i.e., $\|\bm \varphi_{n}\|_{2}=1$ for $n=1,2,\cdots,N$, the condition (\ref{eq7p}) is equivalent to that the Gram matrix $\bm{\Phi}_{\mathcal{K}}^{H}\bm{\Phi}_{\mathcal{K}}$ of every column submatrix $\bm{\Phi}_{\mathcal{K}}(\mathcal{K}\subset \{1,2,\cdots,N\}, |\mathcal{K}|\leq k)$ has all its eigenvalues in the interval $[1-\delta_{k},1+\delta_{k}]$, where $[\ast]^{H}$ denotes the conjugate transpose operation.

Next we give an example in order to clearly illustrate our sampling scheme. If the three undersampled ratios are $p=9,q=10$ and $r=11$, and the number of discrete grid points is $N=100$, the configuration of the samples is
\begin{equation}
\left(
\vphantom{
\begin{array}{*{20}{c}}
{y(9)}&{y(10)}&{y(11)}& \cdots \\
{y(10)}&{y(11)}&{y(18)}& \cdots \\
{y(11)}&{y(18)}&{y(20)}& \cdots \\
{y(18)}&{y(20)}&{y(22)}& \cdots \\
{y(20)}&{y(22)}&{y(27)}& \cdots \\
\vdots & \vdots & \vdots & \ddots
\end{array}
}
\right.
\overbrace{
\begin{array}{*{20}{c}}
{y(9)}&{y(10)}&{y(11)}& \cdots \\
{y(10)}&{y(11)}&{y(18)}& \cdots \\
{y(11)}&{y(18)}&{y(20)}& \cdots \\
{y(18)}&{y(20)}&{y(22)}& \cdots \\
{y(20)}&{y(22)}&{y(27)}& \cdots \\
\vdots & \vdots & \vdots & \ddots
\end{array}
}^{\displaystyle L}
\left.\left.
\vphantom{
\begin{array}{*{20}{c}}
{y(9)}&{y(10)}&{y(11)}& \cdots \\
{y(10)}&{y(11)}&{y(18)}& \cdots \\
{y(11)}&{y(18)}&{y(20)}& \cdots \\
{y(18)}&{y(20)}&{y(22)}& \cdots \\
{y(20)}&{y(22)}&{y(27)}& \cdots \\
\vdots & \vdots & \vdots & \ddots
\end{array}
}
\right)\right\}M.
\end{equation}

When $50$ tasks are used, i.e., $L=50$, it is better to choose $M=27$ to prevent the corresponding sensing matrices from having duplicate rows. The sensing matrix $\bm \Phi_{l}$ amounts to picking partial rows from the $N\times N$ Fourier matrix. The sensing property of this deterministic partial Fourier matrix approximates a random partial Fourier matrix, which has been proven to be appropriate as a CS matrix \cite{candes2006robust}.

\begin{figure}[!htbp]
\centering
\subfigure[$\bm{\Phi}_{1}$]{
\includegraphics[scale=0.7]{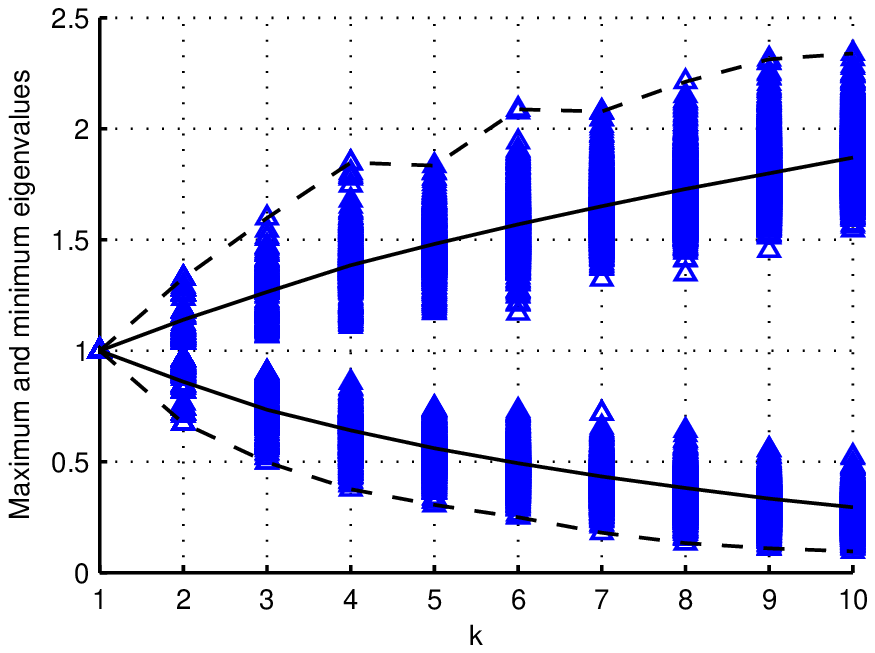}}
\subfigure[The random partial Fourier matrix]{
\includegraphics[scale=0.7]{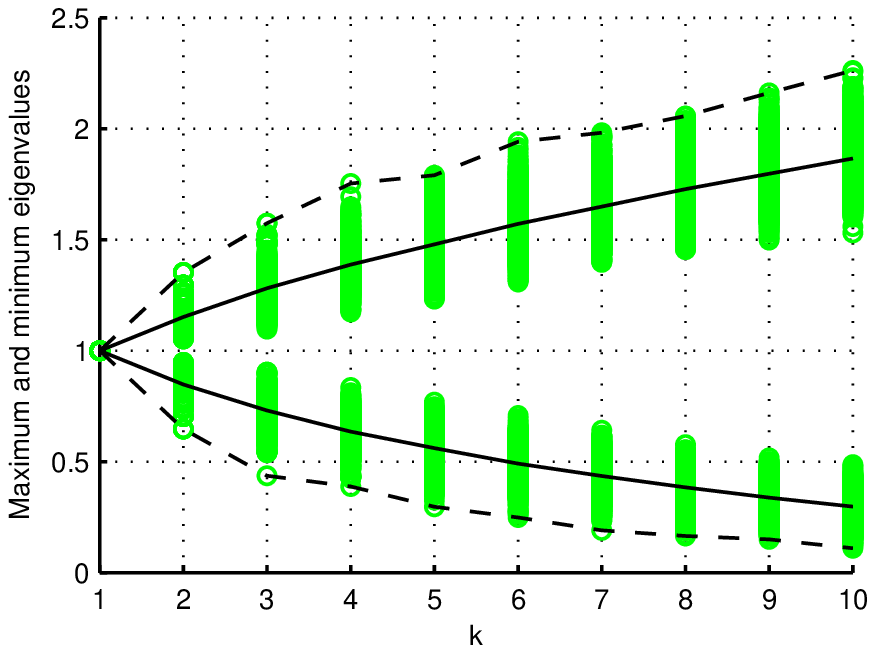}}
\caption{ Maximum and minimum eigenvalues of sub-Gram matrices for different $k$. (a) $\bm{\Phi}_{1}$; (b) The random partial Fourier matrix. } \label{fig2}
\end{figure}

We select $\bm \Phi_{1}$ to present the statistical RIP of $\bm\Phi_{l}$ intuitively, the maximum and minimum eigenvalues of its Gram matrices are plotted. These eigenvalues of a random partial Fourier matrix are also plotted for comparison. The data are obtained from $k^{2}N$ sub-Gram matrices for each $k$. The solid lines sketch the average values of maximum and minimum eigenvalues of all sub-Gram matrices and the dashed lines sketch the limiting values. Fig.~\ref{fig2} shows that the eigenvalues of $\bm{\Phi}_{1}$'s sub-Gram matrices distribute slightly further away from 1 than the random partial Fourier matrix. In Section IV we will see that the probability of success increases significantly when more tasks are introduced.

\section{Proposed Algorithm}
In this section, a complex-valued multitask algorithm based on variational Bayesian inference is proposed to solve the above model. In \cite{ji2009multitask}, the multitask Bayesian CS algorithm utilized empirical Bayesian analysis to recover multiple real-valued sparse solutions. We also address the problem within the hierarchical Bayesian framework. Assume the measurement noise to be independent and complex Gaussian with zero-mean and variance equal to $\beta^{-1}$, the model (\ref{eq6}) can be rewritten as
\begin{equation}\label{eq8}
  {\bm y_l} = {\bm \Phi_l}{\bm s_{l}} + \bm \epsilon_{l} ,~l = 1,2, \cdots ,L.
\end{equation}
The likelihood function for the parameters $\bm s$ and $\beta$ may be expressed as
\begin{equation}\label{eq9}
  p\left( {{\bm y_l}|{\bm s_{l}},\beta } \right) = {\left( {{\pi  \mathord{\left/
 {\vphantom {\pi  \beta }} \right.
 \kern-\nulldelimiterspace} \beta }} \right)^{ - M}}\exp \left( { - \beta \left\| {{\bm y_l} - {\bm \Phi_l}{\bm s_{l}}} \right\|_2^2} \right).
\end{equation}

The hierarchical Gaussian prior is typically imposed on $\bm s_{l}$ in sparse Bayesian leaning to induce sparsity. Denote the prior variance of the $i$-th
element of $\bm s_{l}$ as $\alpha _i^{ - 1}$, the prior distribution of $\bm s_{l}$ is
\begin{equation}\label{eq10}
  p\left( {\bm s_{l}|\bm \alpha } \right) = \frac{1}{{{\pi ^N}{{\left| \bm A  \right|}^{ - 1}}}}\exp \left( { - {\bm s_{l}^H}\bm A \bm s_{l}} \right),
\end{equation}
where $\bm\alpha  = {\left[ {\begin{array}{*{20}{c}}
{{\alpha _1}}& \cdots &{{\alpha _N}}
\end{array}} \right]^T}$ and $\bm A = \text{diag}\left( \bm\alpha  \right)$. Gamma priors are placed on the hyperparameters $\bm \alpha$, and similarly on the noise precision $\beta$, i.e.,
\begin{align}
  \label{eq11} p\left( {\bm \alpha |a,b} \right) &= \prod\limits_{i = 1}^N {\text{Gamma} \left( {{\alpha _i}|a,b} \right)}, \\
  \label{eq12} p\left( {\beta |c,d} \right) &= \text{Gamma}\left( {\beta |c,d} \right).
\end{align}

The parameters $a,b,c$ and $d$ are typically set to very small values (e.g., $a=b=c=d=10^{-6}$), which amounts to assuming uninformative priors for $\bm \alpha$ and $\beta$ \cite{tipping2001sparse}.

Define $\bm Y = \left[ {\begin{array}{*{20}{c}}
{{\bm y_1}}& \cdots &{{\bm y_L}}
\end{array}} \right]$ and $\bm S = \left[ {\begin{array}{*{20}{c}}
{{\bm s_1}}& \cdots &{{\bm s_L}}
\end{array}} \right]$, the joint probability of data, parameters and hyperparameters is
\begin{multline}\label{eq13}
 p\left( {\bm Y,\bm S,\bm \alpha ,\beta } \right) = p\left( {\bm Y|\bm S,\beta } \right)\cdot p\left( {\bm S|\bm \alpha } \right)\cdot p\left( {\bm \alpha |a,b} \right)\cdot p\left( {\beta |c,d} \right)\\
 = \prod\limits_{l = 1}^L {p\left( {{\bm y_l}|{\bm s_l},\beta } \right)}  \cdot \prod\limits_{l = 1}^L {p\left( {{\bm s_l}|\bm \alpha } \right)}  \cdot p\left( {\bm \alpha |a,b} \right) \cdot p\left( {\beta |c,d} \right).
\end{multline}
By applying the variational expectation maximization (EM) algorithm \cite{tzikas2008variational} and the above equations, the posterior distributions of $\bm S,~\bm \alpha$ and $\beta$ can be approximately calculated as
\begin{equation}\label{eq14}
 \ln q(\bm S) = {\left\langle {\ln p\left( {\bm Y,\bm S,\bm \alpha ,\beta } \right)} \right\rangle _{q(\bm \alpha )q(\beta )}} + \text{const},
\end{equation}
\begin{equation}\label{eq15}
  \ln q(\bm \alpha ) = {\left\langle {\ln p\left( {\bm Y,\bm S,\bm \alpha ,\beta } \right)} \right\rangle _{q(\bm S)q(\beta )}} + \text{const},
\end{equation}
and
\begin{equation}\label{eq16}
  \ln q(\beta ) = {\left\langle {\ln p\left( {\bm Y,\bm S,\bm \alpha ,\beta } \right)} \right\rangle _{q(\bm S)q(\bm \alpha )}} + \text{const},
\end{equation}
where $\langle\ast\rangle_{q(x)}$ is the expectation with respect to $q(x)$.

Substituting (\ref{eq13}) into (\ref{eq14}), after some arrangement we find that the vector $\bm s_l$ obeys a complex Gaussian distribution, i.e.,
\begin{equation}\label{eq17}
  q({\bm s_l}) = \mathcal{CN}\left( {{\bm s_l}|{\bm \mu _l},{\bm \Sigma _l}} \right),~l=1,2,\cdots,L.
\end{equation}
The mean $\bm \mu _l$ and covariance matrix ${\bm \Sigma _l}$ are given by
\begin{align}
  \label{eq18} {\bm \mu _l} &= \left\langle \beta  \right\rangle {\bm \Sigma _l}\bm \Phi_l^H{\bm y_l},\\
  \label{eq19} {\bm \Sigma _l} &= {\left( {\left\langle \beta  \right\rangle \bm \Phi _l^H{\bm \Phi _l} + \left\langle \bm A \right\rangle } \right)^{ - 1}}.
\end{align}

According to (\ref{eq13}) and (\ref{eq15}), it can be shown that the posterior density of $\alpha$ is
\begin{equation}\label{eq20}
  q(\bm \alpha ) = \prod\limits_{i = 1}^N {\text{Gamma}\left( {{\alpha _i}|\tilde a,{{\tilde b}_i}} \right)},
\end{equation}
where
\begin{align}
  \label{eq21} \tilde a &= a + L,\\
  \label{eq22} {\tilde b_i} &= b+\left\langle {\sum\limits_{l = 1}^L {{{\left| {{s_{l,i}}} \right|}^2}} } \right\rangle,
\end{align}
$s_{l,i}$ is the $i$-th element of $\bm s_l$. Similarly, we obtain
\begin{equation}\label{eq23}
  q(\beta ) = \text{Gamma}\left( {\beta |\tilde c,\tilde d} \right),
\end{equation}
where
\begin{align}
  \label{eq24} \tilde c &= c + LM,\\
  \label{eq25} \tilde d &= d + \left\langle {\sum\limits_{l = 1}^L {\left\| {{\bm y_l} - {\bm \Phi _l}{\bm s_l}} \right\|_2^2} } \right\rangle.
\end{align}

Utilizing the property of Gamma distribution, the required expected values can be computed as
\begin{align}
  \label{eq26} \left\langle {{\alpha _i}} \right\rangle  &= \frac{{\tilde a}}{{{{\tilde b}_i}}},\\
  \label{eq27} \left\langle \beta  \right\rangle  &= \frac{{\tilde c}}{{\tilde d}}.
\end{align}

Based on the above results, the procedure of the algorithm can be summarized as follows:

1) Set the iteration count to $0$. Initialize $\bm \mu_{l},\bm \Sigma_{l},\bm \alpha$ and $\beta$.

2) According to (\ref{eq20})-(\ref{eq27}) and the current estimated values of $\bm \mu_{l}$ and $\bm \Sigma_{l}$, update the posterior distributions of $\bm \alpha$ and $\beta$.

3) According to (\ref{eq17})-(\ref{eq19}) and the current posterior densities of $\bm \alpha$ and $\beta$, update the the estimated values of $\bm \mu_{l}$ and $\bm \Sigma_{l}$.

4) Return to Step 2) until the iteration count reaches the maximum value.

After using this algorithm, the lowest several valleys of $\bm \alpha$ indicate the positions of the frequencies contained in the signal. When the frequencies do not fall onto the grid, the algorithm often finds the nearest grid point. So the closest interval of the true frequencies can not be too small.

\section{Simulation Results }
\begin{figure}[!htbp]
\centering
\subfigure[$L=1$]{
\includegraphics[scale=0.7]{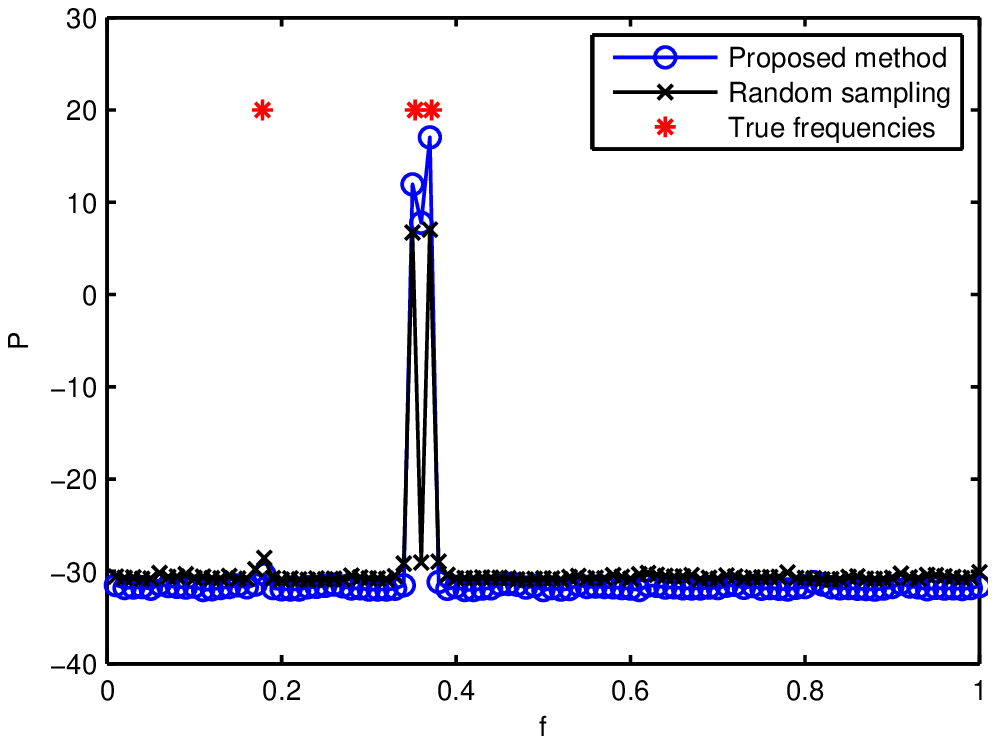}}
\subfigure[$L=10$]{
\includegraphics[scale=0.7]{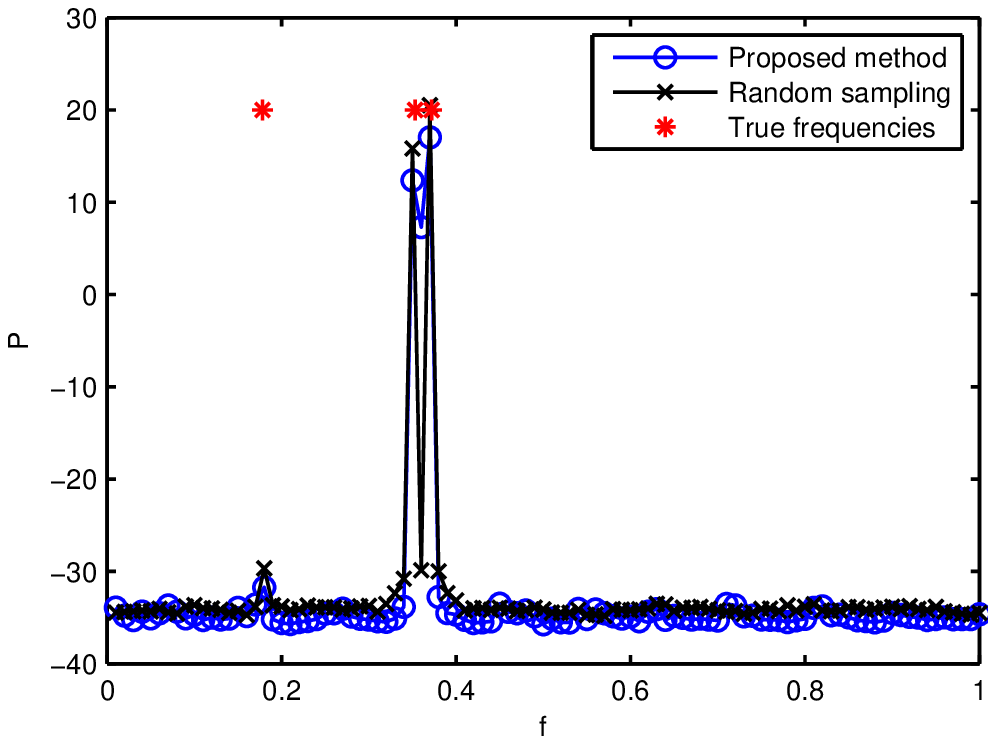}}
\subfigure[$L=30$]{
\includegraphics[scale=0.7]{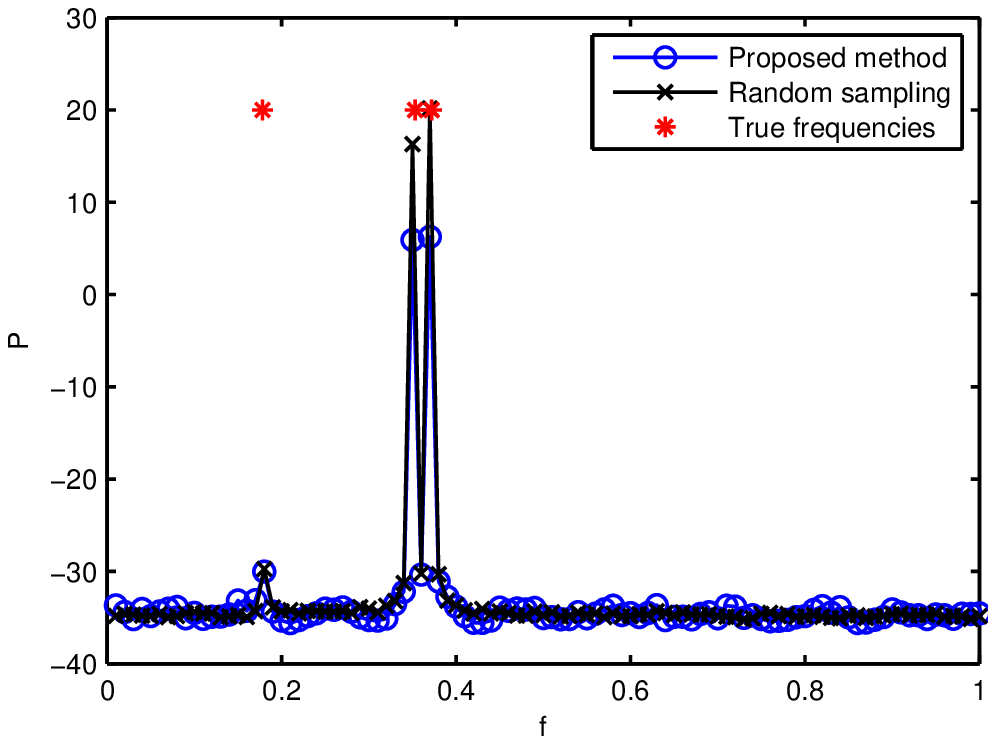}}
\caption{ The estimated power spectra with respect to different numbers of tasks $L$. (a) $L=1$; (b) $L=10$; (c) $L=30$. } \label{fig3}
\end{figure}

We make experiments to verify the effect of multiple tasks relative to single task. The signals contain $K=3$ frequency components with amplitudes of $0.2$, $0.4$ and $0.8$ respectively and random phase angles. The three undersampled ratios are set to $p=9$, $q=10$ and $r=11$. According to the analysis in Section II, $M=27$ and $N=100$ are fixed. We set the frequencies to 0.178, 0.353 and 0.372, respectively. Complex white Gaussian noise at SNR=20dB is added to the measurements. The power spectra with respect to different numbers of tasks $L=1$, 10 and 30 are plotted. Meanwhile, in spite of impracticality of random sampling, we construct multiple random samples in the program to compare with our method. As shown in Fig.~\ref{fig3}, the performance of estimation is improved as the number of tasks $L$. When only one task is utilized, the first frequency component is not really obvious. When $L=1$ and $L=10$, the last two frequency components are not clearly distinguished. When $L=30$, the proposed method achieves the same effects with random sampling.

\begin{figure}[!htbp]
\centering
\includegraphics[scale=0.7]{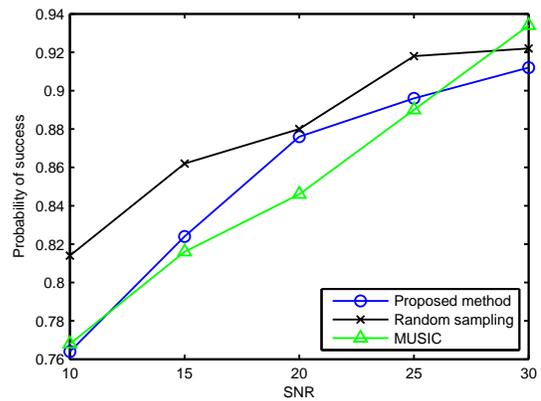}
\centering
\caption{ The probabilities of success for different SNRs. } \label{fig4}
\end{figure}

Then we test the performance of the proposed method in different noisy environment. The three undersampled ratios are set to $p=7$, $q=8$ and $r=9$. $M=32$, $N=100$ and $L=30$ are fixed and the SNR varies from 10dB to 30dB. The signals contain $K=3$ frequency components with random amplitudes and random phase angles. To keep it simple, we assume $K$ is known, so the $K$ frequencies corresponding to maximum $K$ peaks in power spectrum are estimated results. If the deviation of all estimated frequencies from true frequencies are within $0.5/N$, we say this trial is successful. The probabilities of success are obtained from 500 trials for each SNR. We compare the success probabilities of the proposed method with that of random sampling. The MUSIC algorithm using normal sampling is also considered for comparison. The same number of samples are used for our method and MUSIC. As shown in Fig.~\ref{fig4}, the proposed method and MUSIC have approximately the same probabilities of success for different SNRs, which are slightly lower than that of random sampling.

\section{Conclusion}
In this letter, we proposed a deterministic sampling scheme to replace the unpractical random sampling. Three sub-Nyquist analog-to-digital converters which have coprime undersampled ratios are shown to be enough to estimate the spectra of frequency-sparse signals. The property of the corresponding CS matrices is verified through numerical simulations. Then an algorithm based on variational Bayesian inference is proposed to solve the multitask model. Simulations show that this method possesses as good performance as conventional MUSIC with normal sampling. We believe that this method can improve the practicability of CS in line spectral estimation.


%


\ifCLASSOPTIONcaptionsoff
  \newpage
\fi



%
%
%

\bibliographystyle{IEEEtran}
\bibliography{IEEEabrv,mybibfile}

\begin{thebibliography}{10}
\providecommand{\url}[1]{#1}
\csname url@samestyle\endcsname
\providecommand{\newblock}{\relax}
\providecommand{\bibinfo}[2]{#2}
\providecommand{\BIBentrySTDinterwordspacing}{\spaceskip=0pt\relax}
\providecommand{\BIBentryALTinterwordstretchfactor}{4}
\providecommand{\BIBentryALTinterwordspacing}{\spaceskip=\fontdimen2\font plus
\BIBentryALTinterwordstretchfactor\fontdimen3\font minus
  \fontdimen4\font\relax}
\providecommand{\BIBforeignlanguage}[2]{{%
\expandafter\ifx\csname l@#1\endcsname\relax
\typeout{** WARNING: IEEEtran.bst: No hyphenation pattern has been}%
\typeout{** loaded for the language `#1'. Using the pattern for}%
\typeout{** the default language instead.}%
\else
\language=\csname l@#1\endcsname
\fi
#2}}
\providecommand{\BIBdecl}{\relax}
\BIBdecl

\bibitem{donoho2006compressed}
D.~L. Donoho, ``Compressed sensing,'' \emph{IEEE Transactions on Information
  Theory}, vol.~52, no.~4, pp. 1289--1306, 2006.

\bibitem{zhu2011sparsity}
H.~Zhu, G.~Leus, and G.~B. Giannakis, ``Sparsity-cognizant total least-squares
  for perturbed compressive sampling,'' \emph{IEEE Transactions on Signal
  Processing}, vol.~59, no.~5, pp. 2002--2016, 2011.

\bibitem{duarte2013spectral}
M.~F. Duarte and R.~G. Baraniuk, ``Spectral compressive sensing,''
  \emph{Applied and Computational Harmonic Analysis}, vol.~35, no.~1, pp.
  111--129, 2013.

\bibitem{chen2014robust}
Y.~Chen and Y.~Chi, ``Robust spectral compressed sensing via structured matrix
  completion,'' \emph{IEEE Transactions on Information Theory}, vol.~60,
  no.~10, pp. 6576--6601, 2014.

\bibitem{malioutov2005sparse}
D.~Malioutov, M.~{\c{C}}etin, and A.~S. Willsky, ``A sparse signal
  reconstruction perspective for source localization with sensor arrays,''
  \emph{IEEE Transactions on Signal Processing}, vol.~53, no.~8, pp.
  3010--3022, 2005.

\bibitem{bourguignon2007sparsity}
S.~Bourguignon, H.~Carfantan, and J.~Idier, ``A sparsity-based method for the
  estimation of spectral lines from irregularly sampled data,'' \emph{IEEE
  Journal of Selected Topics in Signal Processing}, vol.~1, no.~4, pp.
  575--585, 2007.

\bibitem{chi2011sensitivity}
Y.~Chi, L.~L. Scharf, A.~Pezeshki, and A.~R. Calderbank, ``Sensitivity to basis
  mismatch in compressed sensing,'' \emph{IEEE Transactions on Signal
  Processing}, vol.~59, no.~5, pp. 2182--2195, 2011.

\bibitem{yang2012robustly}
Z.~Yang, C.~Zhang, and L.~Xie, ``Robustly stable signal recovery in compressed
  sensing with structured matrix perturbation,'' \emph{IEEE Transactions on
  Signal Processing}, vol.~60, no.~9, pp. 4658--4671, 2012.

\bibitem{hu2012compressed}
L.~Hu, Z.~Shi, J.~Zhou, and Q.~Fu, ``Compressed sensing of complex sinusoids:
  \text{A}n approach based on dictionary refinement,'' \emph{IEEE Transactions
  on Signal Processing}, vol.~60, no.~7, pp. 3809--3822, 2012.

\bibitem{fang2014super}
J.~Fang, J.~Li, Y.~Shen, H.~Li, and S.~Li, ``Super-resolution compressed
  sensing: \text{A}n iterative reweighted algorithm for joint parameter
  learning and sparse signal recovery,'' \emph{IEEE Signal Processing Letters},
  vol.~21, no.~6, pp. 761--765, 2014.

\bibitem{candes2014towards}
E.~J. Cand{\`e}s and C.~Fernandez-Granda, ``Towards a mathematical theory of
  super-resolution,'' \emph{Communications on Pure and Applied Mathematics},
  vol.~67, no.~6, pp. 906--956, 2014.

\bibitem{tang2013compressed}
G.~Tang, B.~N. Bhaskar, P.~Shah, and B.~Recht, ``Compressed sensing off the
  grid,'' \emph{IEEE Transactions on Information Theory}, vol.~59, no.~11, pp.
  7465--7490, 2013.

\bibitem{yang2015gridless}
Z.~Yang and L.~Xie, ``On gridless sparse methods for line spectral estimation
  from complete and incomplete data,'' \emph{IEEE Transactions on Signal
  Processing}, vol.~63, no.~12, pp. 3139--3153, 2015.

\bibitem{schmidt1986multiple}
R.~O. Schmidt, ``Multiple emitter location and signal parameter estimation,''
  \emph{IEEE Transactions on Antennas and Propagation}, vol.~34, no.~3, pp.
  276--280, 1986.

\bibitem{tropp2010beyond}
J.~A. Tropp, J.~N. Laska, M.~F. Duarte, J.~K. Romberg, and R.~G. Baraniuk,
  ``Beyond nyquist: Efficient sampling of sparse bandlimited signals,''
  \emph{IEEE Transactions on Information Theory}, vol.~56, no.~1, pp. 520--544,
  2010.

\bibitem{candes2008introduction}
E.~J. Cand{\`e}s and M.~B. Wakin, ``An introduction to compressive sampling,''
  \emph{IEEE Signal Processing Magazine}, vol.~25, no.~2, pp. 21--30, 2008.

\bibitem{candes2006robust}
E.~J. Cand{\`e}s, J.~Romberg, and T.~Tao, ``Robust uncertainty principles:
  Exact signal reconstruction from highly incomplete frequency information,''
  \emph{IEEE Transactions on Information Theory}, vol.~52, no.~2, pp. 489--509,
  2006.

\bibitem{ji2009multitask}
S.~Ji, D.~Dunson, and L.~Carin, ``Multitask compressive sensing,'' \emph{IEEE
  Transactions on Signal Processing}, vol.~57, no.~1, pp. 92--106, 2009.

\bibitem{tipping2001sparse}
M.~E. Tipping, ``Sparse \text{B}ayesian learning and the relevance vector
  machine,'' \emph{The journal of machine learning research}, vol.~1, pp.
  211--244, 2001.

\bibitem{tzikas2008variational}
D.~G. Tzikas, A.~C. Likas, and N.~P. Galatsanos, ``The variational
  approximation for \text{B}ayesian inference,'' \emph{IEEE Signal Processing
  Magazine}, vol.~25, no.~6, pp. 131--146, 2008.

\end{thebibliography}
%




\end{document}